\documentclass[a4paper,10pt,twoside]{cpc-hepnp}
\usepackage{multicol}
\usepackage{graphicx}
\usepackage{booktabs}
\usepackage{amssymb,bm,mathrsfs,bbm,amscd}
\usepackage[tbtags]{amsmath}
\usepackage{lastpage}
\usepackage{CJK}
\usepackage{color}
\usepackage{epstopdf}
\usepackage{epsfig}
\usepackage{ulem}
\usepackage{pdfsync}

\begin{document}
\begin{CJK*}{GB}{gbsn}

\title{Running coupling constant at finite chemical potential and magnetic field from holography}

\author{ Xun Chen(陈勋)$^{1,1)}$\email{chenxunhep@qq.com}
\quad Lin Zhang(张林)$^{2,2)}$\email{zhanglin@ucas.ac.cn}
\quad Defu Hou(侯德富)$^{3,3)}$\email{houdf@mail.ccnu.edu.cn}
}
\maketitle

\address{
$^1$ School of Nuclear Science and Technology, University of South China, Hengyang 421001, China\\
$^2$ School of Nuclear Science and Technology, University of Chinese Academy of Sciences,\\Beijing 100049, P.R. China\\
$^3$ Central China Normal University, Wuhan 430079, P.R. China \\}

\begin{abstract}
According to gauge/gravity duality, we use an Einstein-Maxwell-dilaton(EMD) model to study the running coupling constant at finite chemical potential and magnetic field. First, we calculate the effect of temperature on the running coupling constant and find the results are qualitatively consistent with lattice gauge theory. Subsequently, we calculate the effect of chemical potential and magnetic field on running coupling. It is found that the chemical potential and magnetic field both suppress the running coupling constant. However, the effect of the magnetic field is slightly larger than that of chemical potential for a fixed temperature. Compared with the confinement phase, the magnetic field has a large influence on the running coupling in the deconfinement phase.
\end{abstract}

\begin{keyword}
running coulping, holographic QCD, magnetic field
\end{keyword}

\begin{multicols}{2}
\section{Introduction}\label{sec-int}
Color confinement\cite{Weinberg:1973un} and asymptotic freedom\cite{Fritzsch:1973pi,Gross:1973ju,Politzer:1973fx} are important concepts for Quantum Chromodynamics(QCD) as a non-Abelian gauge theory with a coupling constant decreasing with the increase of energy scale. That is, free quarks do not exist, one can only observe color-neutral hadrons at low energy and quarks behave as almost free particles at high momentum transfer\cite{Hebbeker:1992xg}. The investigation of the fundamental forces between quarks and gluons is an essential key to the understanding of QCD and different phases are expected to show up when going from low to high temperatures and/or baryon number densities.

Heavy quarks closely resemble static test charges which can be used to probe microscopic properties of the QCD vacuum, in particular the anatomy of the confinement mechanism\cite{Bali:2000gf}. Forces between static quarks, i.e. static test charges in a thermal and dense medium, change because the gluons, which mediate the interaction between the static quarks, also interact with the constituents (quarks and gluons) of the thermal bath. The interplay between short and long distance length scales plays a crucial role for a quantitative understanding of hard as well as soft processes in relativistic heavy ion collisions\cite{Kaczmarek:2004gv}. As is well known, the running behavior of QCD coupling with densities reflects the essential properties of strongly interacting matter, which can be shown by solving the renormalization group (RG) equation\cite{Li:2016dta}. At short distances precise knowledge of the running coupling constant is needed to match the growing accuracy of hadron scattering experiments as well as to test high-energy models unifying strong and electroweak forces. It is also necessary to know the behavior of thr running coupling constant at long distances, such as the scale of the proton mass, in order to understand hadronic structure, quark confinement and hadronization processes\cite{Deur:2016tte}. In this work, we expect that fundamental forces between quarks and gluons get modified at finite temperature and chemical potential. Furthermore, it has been found in recent years that strong magnetic fields play essential roles in the non-central heavy ion collisions\cite{Kharzeev:2007jp,Kharzeev:2004ey,Fukushima:2008xe,Gatto:2012sp,Abelev:2009ac,Abelev:2009ad,Adamczyk:2013hsi,Adamczyk:2013kcb,Hirano:2010je,Abelev:2012pa,Mo:2013qya,Zhong:2014cda,Zhong:2014sua,Feng:2016srp}. For instance, the magnetic fields produced at the top collision energies of the Large Hadron Collider can reach the order of $eB \sim 15 m_{\pi}^2$. Such strong fields may have consequences for the transport and thermodynamic properties of the quark-gluon plasma(QGP) formed in later stages of heavy ion collisions.

With the discovery of AdS/CFT correspondence\cite{Maldacena:1997re,Witten:1998qj,Gubser:1998bc} or gauge/gravity duality, it has become an effective tool to study the strongly coupled systems. The unique advantage of holography makes it possible for us to study the various properties of the QGP through weakly-coupled gravitational theory. With the development of holographic QCD, the QCD phase transition\cite{Aharony:2006da,Colangelo:2011sr,Evans:2012cx,Evans:2011eu,He:2013qq,Chelabi:2015cwn,Mamo:2015dea,Rougemont:2015oea,Li:2016gfn,Rodrigues:2017iqi,Dudal:2017max,Rodrigues:2018pep,He:2020fdi,Chen:2018vty,Kovensky:2020xif,Rodrigues:2020ndy,Cao:2020ryx,Mamani:2020pks,Ballon-Bayona:2020xls}, glueball and meson spectra\cite{Sakai:2004cn,Karch:2006pv,Colangelo:2008us,Branz:2010ub,Li:2013oda,Braga:2017oqw,Saghebfar:2020ffl,Rinaldi:2020ssz,Zollner:2020nnt}, heavy quark potential\cite{Maldacena:1998im,Andreev:2006eh,Colangelo:2010pe,Li:2011hp,Herzog:2006ra,Ewerz:2016zsx}, jet quenching\cite{Sin:2004yx,Hatta:2008tx,Caceres:2012px,Li:2014hja,Zhu:2019ujc,Zhu:2020wds,Zhang:2020umx} and many aspects\cite{Brodsky:2010ur,Finazzo:2014zga,Rajagopal:2015roa,Braga:2018zlu,DElia:2018xwo,Fukushima:2012kc,Bali:2011qj,Evans:2016jzo,Critelli:2016cvq,
Ballon-Bayona:2017dvv,Gursoy:2016ofp,Gursoy:2017wzz,Fadafan:2015ynz,Chen:2017lsf,Zhang:2016fwr,Zhang:2017izi,Zhang:2020upv,Arefeva:2018hyo,Arefeva:2018cli,Bohra:2019ebj,Zhu:2019igg,Abt:2019tas,Nakas:2020hyo,Elander:2020nyd,Zhou:2020ssi,Saha:2019xbl} of QCD have been studied both from top-down and bottom-up models in the last 20 years.

In this paper, we mainly qualitatively investigate the QCD coupling constant with finite chemical potential and magnetic field at short and large distances in holographic framework. The rest of this paper is organized as follows. In Sec.2, we give a review of the EMD model with two Maxwell fields. In Sec.3, we study the effect of chemical potential and magnetic field on the running coupling constant. Finally, we make a summary in Sec.4.

\section{review of EMD model}\label{02}

First, we briefly review a five-dimensional EMD model with two Maxwell fields\cite{Bohra:2019ebj}. The action of this model is
\begin{gather}
S= \int -\frac{1}{16\pi G_{5}}\sqrt{-g}(R-\frac{f_{1}(\phi)}{4}F_{(1)MN}F^{MN}-\nonumber\\
\frac{f_{2}(\phi)}{4}F_{(2)MN}F^{MN}-\frac{1}{2}\partial_{M}\phi\partial^{M}\phi-V(\phi) )d^{5}x,
\end{gather}
where $F_{(1)MN}$ and $F_{(2)MN}$ are the field strength tensors, $\phi$ is the dilaton field, and $R$ is Ricci scalar. We can consider $A_1$ as the dual of a (neutral) flavor current, capable of creating mesons, while $A_2$ is the dual of the electromagnetic current. $f_{1}(\phi)$, and $f_{2}(\phi)$ are the gauge kinetic functions representing the coupling with the two U(1) gauge fields respectively. $V(\phi)$ is the potential of the dilaton field, whose explicit form will depend on the scale function $A(z)$, and $G_5$ is the Newton constant in five-dimensional spacetime. The presence of magnetic field is introduced by adding a U(1) gauge field which is dual of the electromagnetic current in the gravitational action. For our current purposes, we will just introduce a constant magnetic field $B$, that is, we have no interest in the fluctuations of the Abelian gauge field. We assume the form of metric is
\begin{equation}
ds^{2}=\frac{L^{2} S(z)}{z^{2}}[(-g(z)dt^{2}+\frac{1}{g(z)}dz^{2}+dy^{2}_{1}+e^{B^{2}z^{2}}(dy^{2}_{2}+dy^{2}_{3})],
\end{equation}
where $S(z)$ is the scale factor and we set the AdS radius $L$ to one in this paper. Note that $B$ is the five-dimensional magnetic field which is related to four-dimensional magnetic field with a factor of $\frac{1}{L}$\cite{Dudal:2015wfn}. Thus, the unit of $B$ is GeV. From the action, we can obtain below equations of motion(EoMs)\cite{Bohra:2019ebj}
\begin{equation}
g^{\prime \prime}(z)+g^{\prime}(z)\left(2 B^{2} z+\frac{3 S^{\prime}(z)}{2 S(z)}-\frac{3}{z}\right)-\frac{z^{2} f_{1}(z) A_{t}^{\prime}(z)^{2}}{L^{2} S(z)}=0,
\end{equation}

\begin{gather}
\frac{B^{2} z e^{-2 B^{2} z^{2}} f_{2}(z)}{L^{2} S(z)}+2 B^{2} g^{\prime}(z)\nonumber\\
+g(z)\left(4 B^{4} z+\frac{3 B^{2} S^{\prime}(z)}{S(z)}-\frac{4 B^{2}}{z}\right)=0,
\end{gather}

\begin{equation}
\begin{gathered}
S^{\prime \prime}(z)-\frac{3 S^{\prime}(z)^{2}}{2 S(z)}+\frac{2 S^{\prime}(z)}{z}\\
+S(z)\left(\frac{4 B^{4} z^{2}}{3}+\frac{4 B^{2}}{3}+\frac{1}{3} \phi^{\prime}(z)^{2}\right)=0, \\
\end{gathered}
\end{equation}

\begin{equation}
\begin{gathered}
\frac{g^{\prime \prime}(z)}{3 g(z)}+\frac{S^{\prime \prime}(z)}{S(z)}+S^{\prime}(z)\left(\frac{7 B^{2} z}{2 S(z)}+\frac{3 g^{\prime}(z)}{2 g(z) S(z)}-\frac{6}{z S(z)}\right)+\\
g^{\prime}(z)\left(\frac{5 B^{2} z}{3 g(z)}-\frac{3}{z g(z)}\right)+2 B^{4} z^{2}+\frac{B^{2} z^{2} e^{-2 B^{2} z^{2}} f_{2}(z)}{6 L^{2} g(z) S(z)}\\
-6 B^{2}+\frac{2 L^{2} S(z) V(z)}{3 z^{2} g(z)}+\frac{S^{\prime}(z)^{2}}{2 S(z)^{2}}+\frac{8}{z^{2}}=0,
\end{gathered}
\end{equation}

\begin{equation}
\begin{gathered}
\phi^{\prime \prime}(z)+\phi^{\prime}(z)\left(2 B^{2} z+\frac{g^{\prime}(z)}{g(z)}+\frac{3 S^{\prime}(z)}{2 S(z)}-\frac{3}{z}\right)\\
+ \frac{z^{2} A_{t}^{\prime}(z)^{2}}{2 L^{2} g(z) S(z)} \frac{\partial f_{1}(\phi)}{\partial \phi} \\
-\frac{B^{2} z^{2} e^{-2 B^{2} z^{2}}}{2 L^{2} g(z) S(z)} \frac{\partial f_{2}(\phi)}{\partial \phi}-\frac{L^{2} S(z)}{z^{2} g(z)} \frac{\partial V(\phi)}{\partial \phi}=0,\\
\end{gathered}
\end{equation}

\begin{equation}
A_{t}^{\prime \prime}(z)+A_{t}^{\prime}(z)\left(2 B^{2} z+\frac{f_{1}^{\prime}(z)}{f_{1}(z)}+\frac{S^{\prime}(z)}{2 S(z)}-\frac{1}{z}\right)=0.\\
\end{equation}

We impose the following boundary conditions,

\begin{equation}
\begin{aligned}
&g(0)=1 \text { and } g\left(z_{h}\right)=0, \\
&A_{t}(0)=\mu \text { and } A_{t}\left(z_{h}\right)=0, \\
&S(0)=1, \\
&\phi(0)=0.
\end{aligned}
\end{equation}

Magnetic field is along the $y_1$ direction and we set $S(z) = e^{2 A(z)}$. In this model, $A(e)$ and $f_1(\phi)$ are fixed from QCD phenomena and $V(\phi)$ is solved from the equation of motions. This scenario is named as the potential reconstruction approach introduced in some early literature\cite{Farakos:2009fx,Cai:2012xh,Li:2011hp}. After those, there were many other works exploring this approach in holography. The form of the gauge coupling function $f_1(\phi)$ can be constrained by studying the vector meson mass spectrum. We take the following simple form of $f_{1}(z)=\frac{e^{-c z^{2}-B^{2} z^{2}}}{\sqrt{S(z)}}$ for the reason that the vector meson spectra can be shown to lie on linear Regge trajectories for $B = 0$ and the mass squared of the vector mesons satisfies $m_{n}^{2}=4 c n$. Moreover, the parameter $c$ can also be fixed by matching with the lowest lying heavy meson states $J / \Psi$ and $\Psi'$, and by doing this we get c = 1.16$\rm GeV^2$\cite{Bohra:2019ebj,Dudal:2017max,Yang:2015aia}. The EoMs can be solved as

\begin{gather}
f_{2}(z)=-\frac{L^{2} e^{2 B^{2} z^{2}+2 A(z)}}{z}\left[g(z)\left(4 B^{2} z+6 A^{\prime}(z)-\frac{4}{z}\right)+2 g^{\prime}(z)\right],\\
g(z)=1+ \int_0^{z} d\xi \xi^{3} e^{-B^{2}\xi^{2}-3A(\xi)}(K_{3}+\frac{\widetilde{\mu}^{2}}{2cL^{2}e^{c\xi^{2}}}),\\
K_{3}=-\frac{1+\frac{\widetilde\mu^{2}}{2cL^{2}}\int_0^{z_{h}} d\xi \xi^{3} e^{-B^{2}\xi^{2}-3A(\xi)+c\xi^{2}}}{\int_0^{z_{h}} d\xi \xi^{3} e^{-B^{2}\xi^{2}-3A(\xi)}},\\
\phi(z)=\nonumber\\
\frac{(9a-B^{2})\ln(\sqrt{6a^{2}-B^{4}}\sqrt{6a^{2}z^{2}+9a^{2}-B^{4}z^{2}-B^{2}}+6a^{2}z-B^{4}z)}{\sqrt{6a^{2}-B^{4}}}\nonumber
    \\+z\sqrt{6a^{2}z^{2}+9a-B^{2}(B^{2}z^{2}+1)}-\frac{(9a-B^{2})\log(\sqrt{9a-B^{2}}\sqrt{6a^{2}-B^{4}})}{\sqrt{6a^{2}-B^{4}}},
\end{gather}

\begin{equation}
A_{t}(z)=\mu\left[1-\frac{\int_{0}^{z} d \xi \frac{\xi e^{-B^{2} \xi^{2}}}{f_{1}(\xi) \sqrt{S(\xi)}}}{\int_{0}^{z_{h}} d \xi \frac{\xi e^{-B^{2} \xi^{2}}}{f_{1}(\xi) \sqrt{S(\xi)}}}\right]=\tilde{\mu} \int_{z}^{z_{h}} d \xi \frac{\xi e^{-B^{2} \xi^{2}}}{f_{1}(\xi) \sqrt{S(\xi)}}.
\end{equation}

The metric in the string frame is
\begin{equation}
d s_{s}^{2}=\frac{L^{2} e^{2 A_{s}(z)}}{z^{2}}\left[-g(z) d t^{2}+\frac{d z^{2}}{g(z)}+d y_{1}^{2}+e^{B^{2} z^{2}}\left(d y_{2}^{2}+d y_{3}^{2}\right)\right],
\end{equation}
where $A_{s}=A(z)+\sqrt{\frac{1}{6}}\phi(z)$. Following \cite{Dudal:2017max}, we will assume $A(z)=-a z^{2}$ and take $a = 0.15 \rm{GeV^2}$ for a decent match with the lattice QCD deconfinement temperature at $B$ = 0. The Hawking temperature is given as

\begin{equation}
T=- \frac{z_{h}e^{A(z_{h})-B^{2}z_{h}^{2}}}{4\pi}(K_{3}+\frac{\widetilde{\mu}^{2}}{2cL^{2}}e^{c z_{h}^{2}}).
\end{equation}

The Nambu-Goto action of a string in the worldsheet is given by
\begin{equation}
    S_{NG} = - \frac{1}{2\pi\alpha'} \int d^{2}\xi \sqrt{- \det g_{ab}},
\end{equation}
where $g_{ab}$ is the induced metric, $\frac{1}{2\pi\alpha'}$ is the string tension and
\begin{equation}
    g_{ab} = g_{MN} \partial_a X^M \partial_b X^N, \quad a,\,b=0,\,1.
\end{equation}
Here, $X^M$ and $g_{MN}$ are the coordinates and the metric of the AdS space.

Then, the Nambu-Goto action of the string can be rewritten as

\begin{equation}
    S_{NG} = - \frac{L^2}{2\pi\alpha'T}\int_{-L/2}^{L/2} d{x}^{2} \sqrt{k_1(z) \frac{d{z}^2}{d{x}^2} + k_2(z)}.
\end{equation}

In this case, we can parameterize in the parallel magnetic field direction. $k_1(z)$ and $k_2(z)$ are
\begin{gather}
    k_1(z) =\frac{e^{4A_{s}}}{z^{4}}, \\
    k_2(z) =\frac{e^{4A_{s}}g(z)}{z^{4}}.
\end{gather}

Through the standard procedure, we can get the renormalized two-flavor free energy of the $Q\bar{Q}$ pair,

\begin{equation}
\begin{gathered}
    \frac{\pi F_{Q\bar{Q}}}{\sqrt{\lambda}} = \int_0^{z_0} d{z} (\sqrt{\frac{k_2(z)k_1(z)}{k_2(z)-k_2(z_0)}}-\sqrt{k_2(z\rightarrow 0)}) \\
    - \int_{z_0}^\infty \sqrt{k_2(z\rightarrow 0)} d{z}.
\end{gathered}
\end{equation}

The inter-quark distance of the $Q\bar{Q}$ pair is
\begin{equation}
    r = 2\int_0^{L/2} dx = 2\int_0^{z_0} \frac{dx}{dz} dz =2 \int_0^{z_0} [\frac{k_2(z)}{k_1(z)} (\frac{k_2(z)}{k_2(z_0)}-1)]^{-1/2} d{z}.
\end{equation}

Following the discussions on the running of the QCD coupling\cite{Necco:2001xg,Bali:1992ru,Peter:1997me,Schroder:1998vy,Necco:2001gh}, Refs.\cite{Kaczmarek:2004gv,Kaczmarek:2005ui} define a running coupling constant from the perturbative short and long distance relations for the singlet free energy as

\begin{equation}\alpha_{\mathrm{Q\bar{Q}}}=\frac{3 r^{2}}{4} \frac{\mathrm{d} F_{Q\bar{Q}}}{\mathrm{d} r}.\end{equation}

\section{Numerical results of running coupling constant at finite chemical potential and magnetic field}\label{03}
\begin{figure*}
    \centering
    \includegraphics[width=15cm]{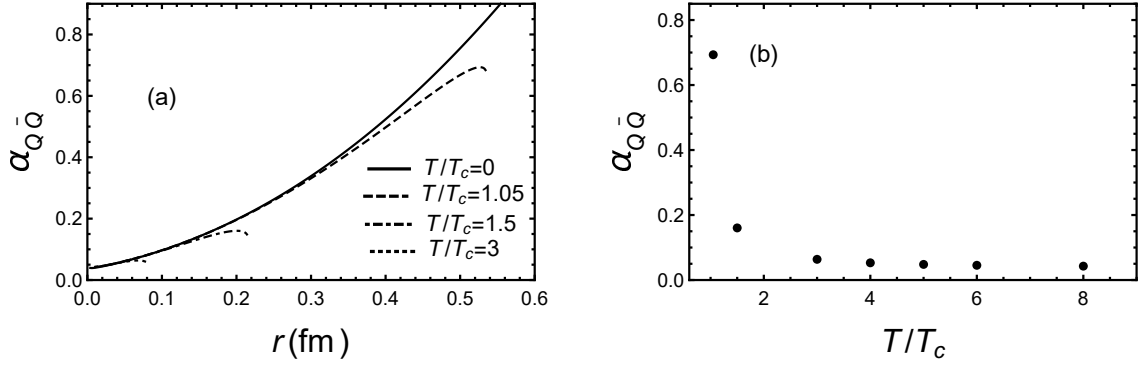}
    \caption{\label{alphatem}(a) Running coupling constant as a function of inter-quark distance L for different temperatures. $T_{c} =266$MeV. Black line is $T/T_{c}=0$, dashed line is $T/T_{c}=1.05$, dot-dashed line is $T/T_{c}=1.5$ and dotted line is $T/T_{c}=3$. } (b) Maximum of running coupling constant at fixed temperature as a function of $T/T_{c}$.
\end{figure*}

Before turning to the finite chemical potential and magnetic field cases, we first numerically calculate the running coupling constant at finite temperature in order to compare with lattice results and other models as shown in Fig.\ref{alphatem}. $T_{c}= 266MeV$ is the deconfinement phase transition temperature of the system at vanishing chemical potential and magnetic field, as determined by free energy\cite{Bohra:2019ebj}.  Fig.\ref{alphatem}(a) shows that the running coupling constant is an increasing function with the increase of inter-quark distance at small $L$. At vanishing temperature, we have $g(z)=1$. For $T/T_{c} = 1.05$, we can see that the running coupling constant changes slowly at small distances($r \leq 0.1$ fm) and then increases quickly until the running coupling reaches a maximum, then decreases quickly. At the same time, the quark-antiquark becomes deconfined. For higher temperature, the maximum running coupling decreases. From the above discussion, we know that the temperature will significantly affect the running coupling constant at large distances. However, the temperature has little impact on the running coupling constant at the small distance. We take the maximum running coupling constant at fixed temperature and plot in Fig.\ref{alphatem}(b). This shows that the temperature will significantly influence the maximum value of running coupling constant and it will slowly tend to zero at extremely high temperatures. These results are qualitatively consistent with lattice\cite{Kaczmarek:2004gv} and other works\cite{Ewerz:2013sma,Ewerz:2015jno} which use the original soft-wall qualitative analysis in this paper. Nevertheless, the EMD model is self-consistently solved from the Einstein field equations and the EoMs for $\phi$ and $A_t$. If we want to get the results very close to lattice, we need to adjust the form of $A(z)$ and the parameters. We only focus on qualitative analysis in this paper.

\begin{figure*}
    \centering
    \includegraphics[width=15cm]{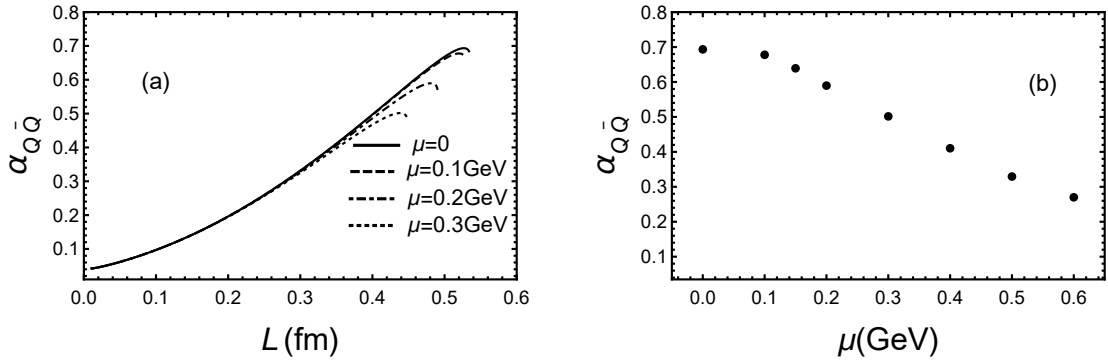}
    \caption{\label{alphache}(a) Running coupling constant as a function of inter-quark distance L for different chemical potentials for a fixed temperature $T = 1.05T_{c}$. Black line is $\mu = 0$, dashed line is $\mu = 0.1GeV$, dot-dashed line is $\mu = 0.2GeV$ and dotted line is $\mu = 0.3GeV$. (b) Maximum of running coupling constant at fixed temperature as a function of chemical potential $\mu$.}
\end{figure*}

In Fig.\ref{alphache}, we show the effect of chemical potential on the running coupling constant for a fixed temperature $T = 1.05T_{c}$. We found that the maximum running coupling is a decreasing function of chemical potential and the chemical potential still has little impact on the running coupling constant at small distances. From the right panel of the maximum running coupling decreases slowly at small chemical potential then decreases quickly at large chemical potential. However, the effect of chemical potential on the running coupling constant is still less dramatic than the effect of temperature. The qualitative results are in accordance with the quasiparticle model\cite{Zheng:2005rtn,Lu:2016fki}.

\end{multicols}

\begin{figure*}
    \centering
    \includegraphics[width=15cm]{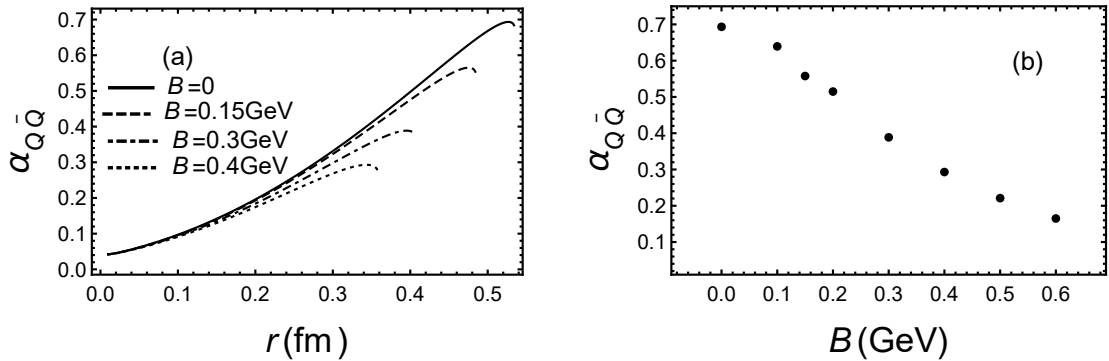}
    \caption{\label{alphamag}(a) Running coupling constant as a function of inter-quark distance L for different magnetic fields in a fixed temperature $T = 1.05T_{c}$. Black line is $B=0$, dashed line is $B = 0.15GeV$, dot-dashed line is $B = 0.3GeV$ and dotted line is $B = 0.4GeV$. (b) Maximum of running coupling constant in fixed temperature as a function of magnetic field $B$.   }
\end{figure*}

\begin{figure*}[h]
    \centering
    \includegraphics[width=8cm]{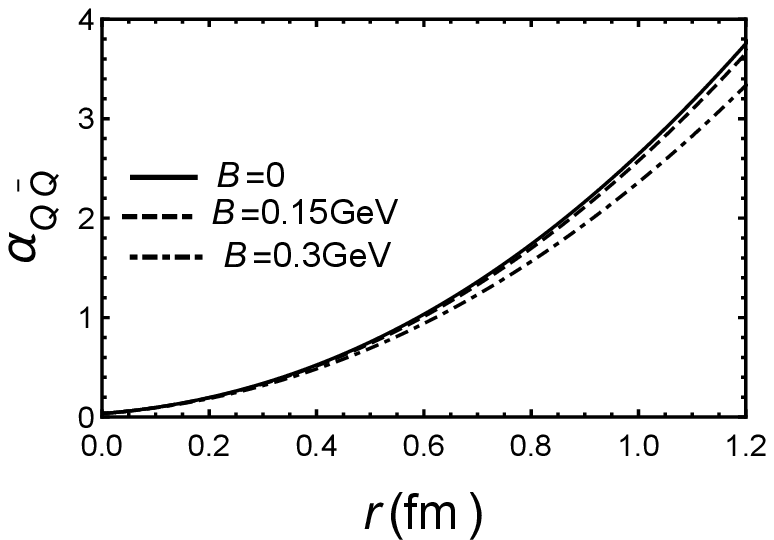}
    \caption{\label{confine} Running coupling constant as a function of inter-quark distance L for different magnetic fields. Black line is $B=0$, dashed line is $B = 0.15GeV$, dot-dashed line is $B = 0.3GeV$.   }
\end{figure*}

\begin{multicols}{2}
Next, we show the effect of parallel magnetic field on the running coupling constant for a fixed temperature $T = 1.05T_{c}$ in Fig.\ref{alphamag}. The qualitative behavior of the running coupling constant under the magnetic field is similar to that at finite temperature and chemical potential. But the effect of magnetic field is larger than chemical potential and smaller than temperature. From the right panel of Fig.\ref{alphamag}(b), we can see the maximum running coupling constant will decrease smoothly with magnetic field. The direction of magnetic field perpendicular to the string can also be calculated in principle, but we haven't shown the results here since we find the difference between parallel and perpendicular direction is extremely small for the magnetic field created in present experiments. The Polyakov-Nambu-Jona-Lasinio(PNJL) model\cite{Farias:2014eca,Li:2016dta,Ferreira:2014kpa} also shows that the running coupling constant decreases with the magnetic field strength, which is also consistent with our model.

Finally, we discuss the running coupling in the confined phase which corresponds to thermal-AdS in this model. It is easy to find that the metric is independent of $z_h$ and $\mu$.  In the confined phase, the inter-quark distance can't be infinite as discussed in Ref.~\cite{Andreev:2021bfg,Chen:2021bkc}. The distance of string breaking occurs at around $L = 1.2\rm{fm}$ and a heavy meson decays to a pair of heavy-light mesons. The detailed discussion of string breaking in different magnetic fields will be left to future work. We can study the effect of magnetic field on the running coupling in Fig.\ref{confine}. We found that the running coupling will decrease with the increase of magnetic field and the influence of magnetic field in the deconfined phase is larger than the influence of magnetic field in the confined phase.
\end{multicols}
\begin{multicols}{2}
\section{Conclusion}\label{04}

In this paper, we use a holographic model to calculate the running coupling constant constant at finite chemical potential and magnetic field. First, we calculate the effect of temperature on running coupling constant constant. In the Fig. 2 of Ref. \cite{Kaczmarek:2004gv}, lattice results show that $\alpha_{qq}$ agrees with the zero temperature result at short distance. Thermal effects only become visible at larger distances and lead to a decrease of the coupling relative to its zero temperature value. At the temperature $T>T_c$, running coupling has a maximum value beyond which the running coupling will decrease with the increase of distance and the maximum value decreases with the increase of temperature. The results of this holographic model also are qualitatively consistent with lattice results in Fig. \ref{alphatem}.

Then, we study the chemical potential and magnetic field on running coupling. The interaction between two quarks are reflected by the running coupling or the heavy quark-antiquark potential(free energy). In our previous study\cite{Zhou:2020ssi,Chen:2017lsf}, it is found that chemical potential and magnetic field will change the free energy of $Q\bar{Q}$. Since the running coupling is extracted from the free energy of $Q\bar{Q}$\cite{Kaczmarek:2004gv,Kaczmarek:2005ui}, we can infer that the chemical potential and magnetic field should have an influence on the running coupling. More discussions of the running coupling in the presence of the chemical potential and magnetic field can be found in Ref.\cite{Zheng:2005rtn,Lu:2016fki,Farias:2014eca,Li:2016dta,Ferreira:2014kpa}. In our paper, we find the presence of chemical potential and magnetic field will decrease the maximum of running coupling constant and have little on influence running coupling at small distance(large momentum transfer). Besides, The influence of the magnetic field on running coupling constant is larger than the chemical potential for a fixed temperature. Finally, we find the magnetic field has large influence on the running coupling in the deconfinement phase compared with the confinement phase.

\section{Acknowledgments}

\acknowledgments{X.C. is supported by the Research Foundation of Education Bureau of Hunan Province, China (Grant No. 21B0402). L.Z. is supported in part by the Fundamental Research Funds for the Central Universities. D.H. is in part supported by the NSFC Grant Nos. 11735007, 11890711.}

\end{multicols}

\vspace{10mm}

\begin{multicols}{2}

\end{multicols}

\vspace{-1mm}
\centerline{\rule{80mm}{0.1pt}}
\vspace{2mm}

\begin{multicols}{2}

\end{multicols}

\clearpage

\end{CJK*}
\end{document}